\documentclass{article}

\usepackage{amsmath,graphicx,subfig}
\usepackage{amssymb,amsfonts}
\usepackage{overpic}
\usepackage[cal=boondox]{mathalpha}
\usepackage{bm}
\usepackage{enumitem}
\graphicspath{ {./images/} }
\usepackage{multirow}
\usepackage{array}
\usepackage{mdwmath}
\usepackage{mdwtab}
\usepackage{booktabs}
\usepackage[nodisplayskipstretch]{setspace} 
\usepackage[ruled,vlined,lined,commentsnumbered]{algorithm2e}
\usepackage[compact]{titlesec}
\usepackage{arxiv}

\usepackage[utf8]{inputenc} % allow utf-8 input
\usepackage[T1]{fontenc}    % use 8-bit T1 fonts
\usepackage{hyperref}       % hyperlinks
\usepackage{url}            % simple URL typesetting
\usepackage{booktabs}       % professional-quality tables
\usepackage{amsfonts}       % blackboard math symbols
\usepackage{nicefrac}       % compact symbols for 1/2, etc.
\usepackage{microtype}      % microtypography
\usepackage{lipsum}		% Can be removed after putting your text content
\usepackage{graphicx}
\usepackage{doi}
\usepackage{csquotes}

\usepackage{tikz}
\usepackage{textcomp}

\title{Frequency Estimation Using Complex-Valued Shifted Window Transformer}

\author{\href{https://orcid.org/0000-0002-3388-4805}{\includegraphics[scale=0.06]{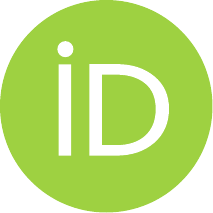}\hspace{1mm}Josiah W. Smith} \\
	Department of Electrical and Computer Engineering\\
	The University of Texas at Dallas\\
	Richardson, TX 75080 \\
	\texttt{josiah.smith@utdallas.edu} \\
	\And
	\href{https://orcid.org/0000-0001-7229-1765}{\includegraphics[scale=0.06]{orcid.pdf}\hspace{1mm}Murat Torlak} \\
	Department of Electrical and Computer Engineering\\
	The University of Texas at Dallas\\
	Richardson, TX 75080 \\
	\texttt{torlak@utdallas.edu} 
}

\date{}

\renewcommand{\shorttitle}{Frequency Estimation Using Complex-Valued Shifted Window Transformer}

\hypersetup{
pdftitle={Frequency Estimation Using Complex-Valued Shifted Window Transformer},
pdfsubject={cs.AI, eess.SP},
pdfauthor={Josiah W.~Smith},
pdfkeywords={Complex-valued neural network (CVNN), deep learning, frequency estimation, radar, spectral super-resolution, Swin transformer},
}

\begin{document}
\maketitle

\begin{abstract}
Estimating closely Estimating closely spaced frequency components of a signal is a fundamental problem in statistical signal processing. 
In this letter, we introduce 1-D real-valued and complex-valued shifted window (Swin) transformers, referred to as \textit{SwinFreq} and \textit{CVSwinFreq}, respectively, for line-spectra frequency estimation on 1-D complex-valued signals. 
Whereas 2-D Swin transformer-based models have gained traction for optical image super-resolution, we introduce for the first time a complex-valued Swin module designed to leverage the complex-valued nature of signals for a wide array of applications.  
The proposed approach overcomes the limitations of the classical algorithms such as the periodogram, MUSIC, and OMP in addition to state-of-the-art deep learning approach cResFreq.
SwinFreq and CVSwinFreq boast superior performance at low signal-to-noise ratio SNR and improved resolution capability while requiring fewer model parameters than cResFreq, thus deeming it more suitable for edge and mobile applications. 
We find that the real-valued Swin-Freq outperforms its complex-valued counterpart CVSwinFreq for several tasks while touting a smaller model size. 
Finally, we apply the proposed techniques for radar range profile super-resolution using real data. 
The results from both synthetic and real experimentation validate the numerical and empirical superiority of SwinFreq and CVSwinFreq to the state-of-the-art deep learning-based techniques and traditional frequency estimation algorithms. 
The code and models are publicly available at \href{https://github.com/josiahwsmith10/spectral-super-resolution-swin}{https://github.com/josiahwsmith10/spectral-super-resolution-swin}.
\end{abstract}

% keywords can be removed
\keywords{Complex-valued neural network (CVNN) \and deep learning \and frequency estimation \and radar \and spectral super-resolution \and Swin transformer.}

\section{Introduction}
For signal processing applications such as remote sensing, radar imaging, communications, localization, sonar, etc. \cite{izacard2019aLearning,izacard2021datadriven,pan2021deep,pan2021complexFrequencyEstimation}, estimating the frequency content of multisinusoidal signal is an essential aspect of robust sensing, ranging, target recognition, or navigation. 
Many radar systems, for example, model the demodulated signal as a multisinusoidal signal whose unique frequency components correspond to scatterers at different ranges or directions and targets in the sensor field of view (FoV). 
The classical periodogram, which is often implemented as a windowed Fourier transform, is commonly used for frequency estimation, but suffers from limited resolution due to mutual interference and the \textit{sinc}-effect, causing it to fail when frequency components are closely located, particularly for weak components that are often subjugated by an adjacent strong component. 
Research over recent decades has yielded impressive progress towards subspace-based techniques and sparse representation-based algorithms to surpass the Rayleigh limit. 
Notably, the MUltiple SIgnal Classification (MUSIC) and estimation of signal parameters via rotational invariance techniques (ESPRIT) algorithms yield satisfactory frequency resolution under sufficient signal-to-noise (SNR) conditions. 
However, as SNR declines to real-world levels, performance degrades substantially. 
Alternatively, compressed sensing (CS) theory has produced algorithms such as orthogonal matching pursuit (OMP), a classical greedy algorithm, to solve the sparse representation problem for frequency estimation. 
However, MUSIC, ESPRIT, and OMP require the number of target reflectors in the scene as \textit{a priori} for defining signal subspace decomposition. 

More recently, deep learning-based algorithms, which have shown tremendous potential on optical images and natural language, has been adopted for statistical signal processing applications, including frequency estimation. 
Data-driven frequency estimation algorithms have been introduced for 1-D \cite{izacard2019aLearning,izacard2021datadriven} and 2-D \cite{pan2021deep} multisinusoidal signals, referred to as DeepFreq and ResFreq.
However, the magnitude and phase relationships of the complex-valued signals are ignored as the entire model relies on real-valued computations, resulting in degraded performance. 
Later, the authors of \cite{pan2021deep} examined the line-spectra frequency estimation problem to produce the state-of-the-art cResFreq algorithm, which replaces the real-valued matched filter (MF) module of ResFreq with a complex-valued alternative \cite{pan2021complexFrequencyEstimation}. 
However, after the MF module, the signal modulus is computed and the second half of the neural network, dubbed the super-resolution (SR) module, employs only real-valued neural network operations. 
In this letter, we propose a frequency estimation network with a similar construction to cResFreq, being composed of a MF module followed by a SR module; however, our model employs both a complex-valued MF module and complex-valued SR module. 

Although complex-valued neural networks (CVNNs) have been explored in previous literature for complex-valued multi-layer perceptron and convolutional neural networks \cite{kim1990modification,lee2022complex,smith2023dual_radar}, we introduce a complex-valued deep learning module based on the popular shifted window (Swin) transformer \cite{liu2021swin}. 
Leveraging the attention mechanism and vision transformer (ViT) \cite{dosovitskiy2020image_ViT}, the Swin transformer reduces the computational complexity from quadratic to linear by limiting self attention to non-overlapping local windows. 
Inspired by the traditional convolution, consecutive Swin layers shift the window partition providing connections that significantly improve performance \cite{liu2021swin}. 
To the authors' knowledge, this letter details the first implementation of a complex-valued Swin transformer or 1-D Swin transformer suitable for application to many statistical signal processing problems that rely on vector data. 
We implement both the the real-valued Swin-based frequency estimation network, referred to as \textit{SwinFreq}, and its complex-valued counterpart, \textit{CVSwinFreq}.  
Using the proposed real-valued and complex-valued Swin layer, we introduce a novel residual signal Swin transformer block, a sequence of which are used to construct the SR module. 
The proposed algorithm is validated by synthetic and real data to outperform the classical periodogram, MUSIC, OFDM, and cResFreq \cite{pan2021complexFrequencyEstimation}, in terms of spectral reconstruction fidelity, sidelobe reduction, and resolution capability thereby establishing a new state-of-the-art for line-spectra super-resolution. 

In summary, the contributions of this letter are as follows. 
\begin{enumerate}
    \item[1)] We introduce novel real-valued and complex-valued shifted window (Swin) transformer models, SwinFreq and CVSwinFreq, suitable for application to a wide array of data-driven signal processing problems. 
    \item[2)] Whereas \cite{pan2021complexFrequencyEstimation} employs a partially complex-valued approach, we introduce the first frequency estimation network to employ a fully-complex-valued neural network model. 
    \item[3)] The proposed 1-D Swin transformer architectures achieve superior numerical performance compared with cResFreq while boasting smaller model sizes and hence smaller memory demands for practical implementations.  
    However, although CVSwinFreq outperforms SwinFreq and cResFreq for resolution of closely spaced frequencies, the real-valued SwinFreq achieves better performance at most other tasks while requiring fewer model parameters. 
    \item[4)] The proposed algorithms are applied to real radar imaging data demonstrating robustness for practical implementations.
\end{enumerate}
The rest of this letter is organized as follows. 
Section \ref{sec:methods} introduces the frequency estimation problem and details the proposed techniques. 
Section \ref{sec:results} overviews experimental results, followed by conclusions. 

\section{Frequency Estimation using CVNNs}
\label{sec:methods}

%in the ideal noiseless case can be expressed as
%\begin{equation}
%    \label{eq:multisinusoidal}
%    s(t) = \sum_{\ell = 0}^{L - 1} \alpha_\ell e^{j 2 \pi f_\ell t}, 
%\end{equation}
%where $L$ is the number of frequencies and $t$ represents the axis over which the multisinusoidal signal is measured, which could be time, frequency, space, etc. depending on the application. 
%For the ideal case in \eqref{eq:multisinusoidal}, the amplitudes, $\alpha_\ell \in \mathbb{C}$, and respective frequencies, $f_\ell \in [-0.5, 0.5)$, can be perfectly recovered. 
%However, for practical implementations, additional noise and sampling constraints degrade the frequency estimation performance. 
%For any practical system, only a finite number of samples can be collected, imposing a windowing function on the collected signal. 
%Together with an additive noise term, the practical multisinusoidal signal after an analog-to-digital converter (ADC) can be expressed as 
%where $T_s$ is the sampling period, $T_c$ is the collection period during which all samples are taken,  $\Pi(t)$ is the rectangular window function 

Across a wide range of signal processing problems, the frequency estimation problem is a crucial issue for multisinusoidal signals. 
The general form of a sampled multisinusoidal signal under an additive noise model can be expressed as
\begin{equation}
    \label{eq:practical_multisinusoidal}
    s[n] = \sum_{\ell = 0}^{L - 1} \alpha_\ell e^{j 2 \pi f_\ell n} + z[n], \quad n = 0,1,\dots,N-1,
\end{equation}
where $L$ is the number of frequency components, $f_\ell \in [-0.5, 0.5)$ are the digital frequencies whose corresponding amplitudes are $\alpha_\ell \in \mathbb{C}$, $N$ is the number of samples, and $z[n]$ is the additive noise. 
In the ideal case, without noise and where an infinite number of samples can be collected, the frequencies components can be exactly recovered. 
However, for any practical case following \eqref{eq:practical_multisinusoidal}, additional noise and sampling constraints degrade the frequency estimation performance. 
Limiting the samples between $0$ and $N-1$ effectively imposes a rectangular window on the signal $s[n]$. 
The multiplication of the rectangular window function corresponds to convolution of a discrete sinc or Dirichlet kernel in the frequency domain of width $1/N$, known as the sinc-effect. 
This results in a blurring of the spectral information of the signal, together along with the addition of undesirable sidelobe-effects resulting in overlapping and distortion of the ideal frequency components, as illustrated in \cite{izacard2021datadriven}. 
Hence, the frequency estimation problem is also referred to as \textit{spectral super-resolution}, wherein the objective is to super-resolve the spectral components of a signal \cite{izacard2019aLearning}. 

Hence, spectral estimation and super-resolution has garnered considerable attention from the research community in recent decades. 
A natural solution to the frequency estimation problem is to employ a matched filter response, often referred to as the \textit{periodogram}. 
The periodogram correlates the signal with a set of kernels centered at known frequencies, and is commonly computed using a windowed discrete Fourier transform \cite{pan2021complexFrequencyEstimation}. 
However, although windowing can help mitigate sidelobe-effects, it further degrades resolution performance for closely spaced frequency components \cite{izacard2021datadriven}. 
Taking a different approach, by assuming a small $L$ with respect to $N$, the popular MUltiple SIgnal Classification (MUSIC) algorithm leverages the low rank of the sample covariance matrix. 
Using eigendecomposition, MUSIC separates the signal and noise subspaces to compute a super-resolved line-spectra. 
However, the performance of MUSIC significantly degrades at low SNR, adding to its implementation challenge of relatively high computational complexity. 
Alternatively, sparse representation algorithms have shown impressive results for spectral super-resolution. 
The orthogonal matching pursuit (OMP) algorithm is a greedy approach that iteratively computes locally optimal nonsparse coefficients to efficiently generate a solution that promises local optimality. 
However, OMP estimates become biased for closely spaced frequencies or low SNRs. 
Additionally, both MUSIC and OMP require \textit{a priori} knowledge of the number of frequency components. 
To address the latter issue, we apply a classical information theoretic criterion such as second-order statistic of eigenvalues (SORTE) or Akaike information criterion (AIC) \cite{izacard2021datadriven}. 

\begin{figure}[ht]
    \centering
    \includegraphics[width=0.65\textwidth]{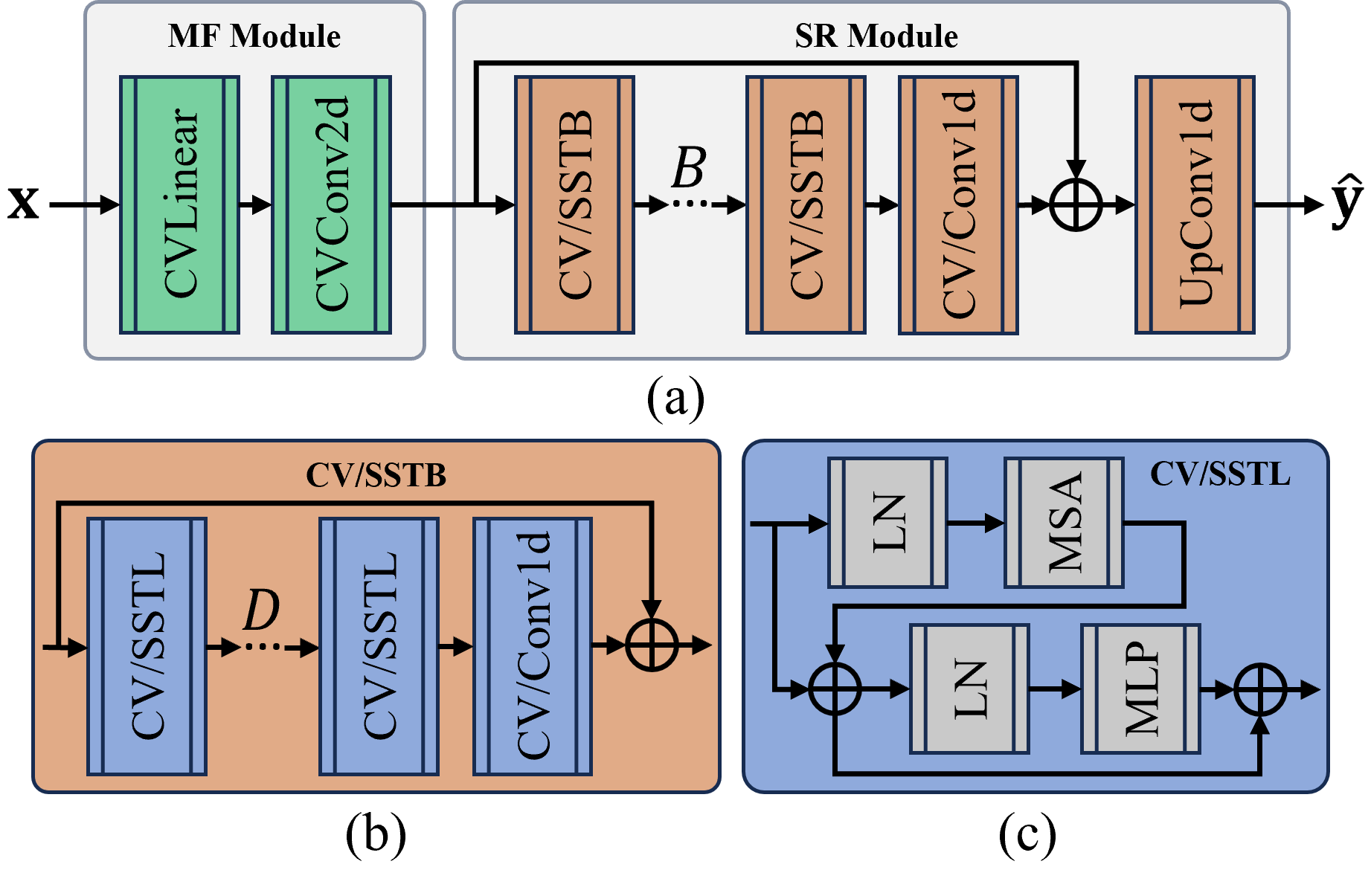}
    \caption{Architecture of the (a) proposed SwinFreq or CVSwinFreq model, (b) SSTB or CVSSTB module, and (c) SSTL or CVSSTL module.}
    \label{fig:model_arch}
\end{figure}

To address the deficiencies of the periodogram, MUSIC, and OMP, deep learning approaches recently been introduced such as DeepFreq \cite{izacard2019aLearning,izacard2021datadriven}, ResFreq \cite{pan2021deep} and state-of-the-art cResFreq \cite{pan2021complexFrequencyEstimation} for frequency estimation. 
Our proposed algorithm adopts a similar structure to cResFreq being composed of two primary components: the matched filter (MF) module and super-resolution (SR) module, as shown in Fig. \ref{fig:model_arch}. 
The MF module and transposed convolution ($\mathrm{UpConv1d}$) employ identical architectures to cResFreq with the exception of a more efficient computation of the complex-valued linear and convolution layers using Gauss' multiplication trick and more robust complex-valued weight initialization strategies \cite{smith2023complex}. 
Given a complex-valued time-series sample $\mathbf{X} \in \mathbb{C}^N$, the MF module, whose weight vectors are shown to be close to a set of complex exponential functions \cite{pan2021complexFrequencyEstimation}, yields the spectral feature $\mathbf{F}_0 \in \mathbb{C}^{M \times C}$ as
\begin{equation}
    \label{eq:mf_module}
    \mathbf{F}_0 = \mathsf{H}_\mathrm{MF}(\mathbf{X}) = \mathsf{H}_\mathrm{CVConv2d}\left( \mathsf{H}_\mathrm{CVLinear}(\mathbf{X}) \right).
\end{equation}
However, our SR module has two notable differences. 
First, similar to \cite{liang2021swinir,smith2023dual_radar}, we incorporate a residual connection to pass forward the intermediate features extracted by the MF module. %, which is added to the result of the sequence of \textit{Signal Swin Transformer Blocks (SSTBs)}. 
Additionally, rather than relying on a convolution as the backbone for the SR module, we introduce a new module based on the Swin transformer called the \textit{Complex-Valued Signal Swin Transformer Block (CVSSTB)} and its real-valued counterpart, SSTB, as shown in Fig. \ref{fig:model_arch}b. 
Provided the spectral feature $\mathbf{F}_0$ from the MF module, the SR module reconstructs the super-resolved spectrum of the input signal by first applying a series of CVSSTBs or SSTBs followed by a transposed convolution layer for high-quality frequency-domain reconstruction. 
More specifically, the intermediate features of the SR module, $\mathbf{F}_1,\mathbf{F}_2,\dots,\mathbf{F}_B \in \mathbb{C}^{M \times C}$, and output, $\hat{\mathbf{Y}} \in \mathbb{C}^{N_{SR}}$ are computed as
\begin{equation}
\begin{gathered}
    \label{eq:sr_module}
    \mathbf{F}_i = \mathsf{H}_{\mathrm{CV/SSTB}_i}(\mathbf{F}_{i-1}), \quad i = 1,2,\dots,B, \\
    \hat{\mathbf{Y}} = \mathsf{H}_\mathrm{UpConv1d}(\mathbf{F}_0 + \mathbf{F}_B). 
\end{gathered}
\end{equation}
%where $H_{CV/SSTB_i}(\cdot)$ denotes the $i$-th CVSSTB or SSTB and $H_\text{UpConv1d}(\cdot)$ is the output layer. 
Each CVSSTB or SSTB applies a series of \textit{Complex-Valued Signal Swin Transformer Layers (CVSSTLs)} or real-valued SSTLs, respectively, and can be computed as
\begin{equation}
\begin{gathered}
    \label{eq:cvsstb}
    \mathbf{G}_{k,i} = \mathsf{H}_{\mathrm{CV/SSTL}_k}(\mathbf{G}_{k-1,i}), \quad k = 1,2,\dots,D, \\
    \mathbf{F}_i = \mathsf{H}_{\mathrm{CV/Conv1d}_i}(\mathbf{G}_{0,i} + \mathbf{G}_{D,i}), 
\end{gathered}
\end{equation}
where $D$ is the number of CVSSTLs or SSTLs, $\mathsf{H}_{\mathrm{CV/Conv1d}_i}$ denotes the complex-valued or real-valued 1-D convolution layer of the $i$-th CVSSTB or SSTB, and $\mathbf{G}_{0,i} = \mathbf{F}_{i-1}$. 

\begin{figure}[ht]
    \centering
    \includegraphics[width=0.65\textwidth]{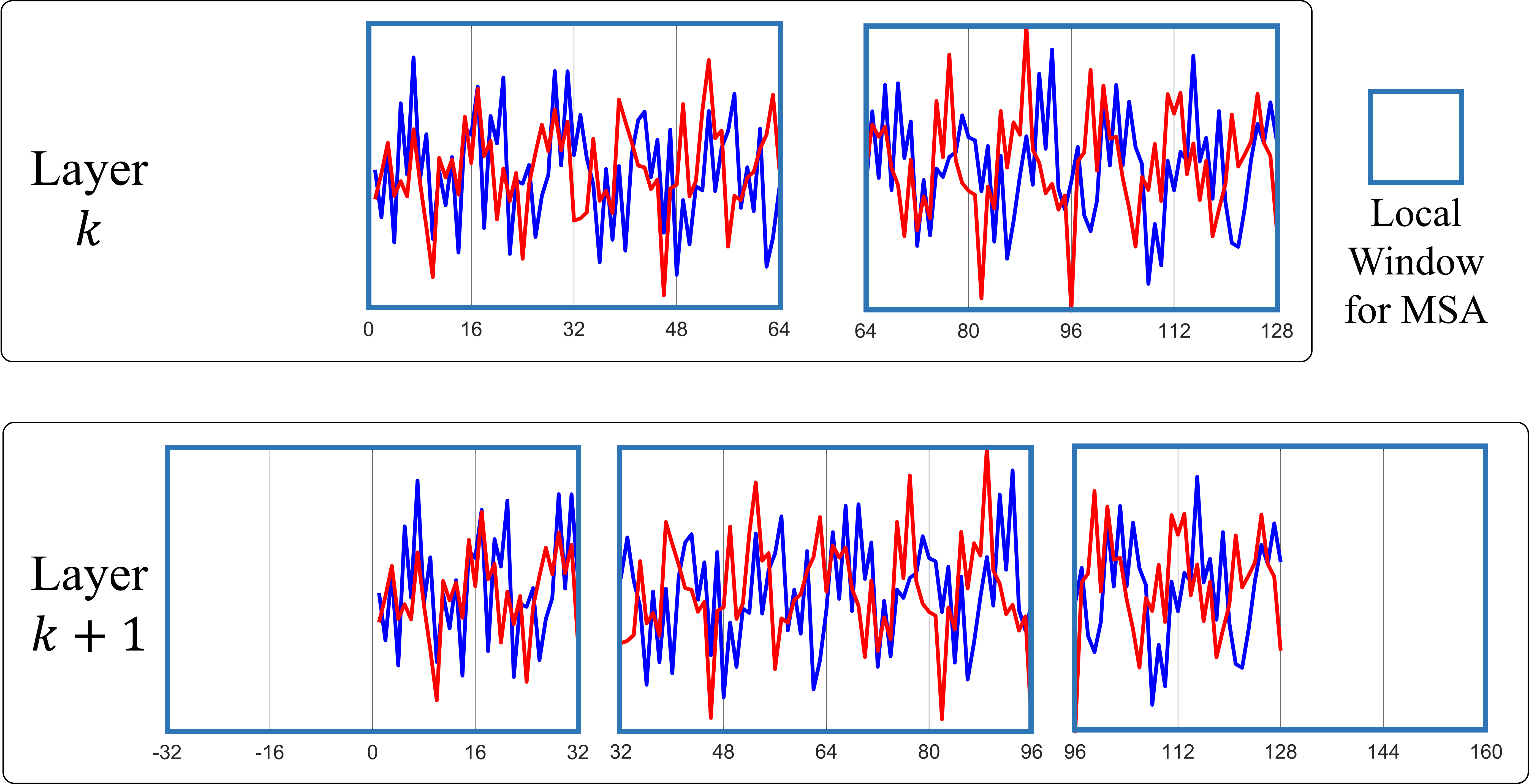}
    \caption{Illustration of shifted window configuration for consecutive layers. In the $k$-th layer, the complex-valued signal -- whose real and imaginary parts are represented in \textcolor{red}{red} and \textcolor{blue}{blue} -- is partitioned using the standard windowing scheme. In subsequent layer ($k+1$)-th, the window partition is shifted. By computing self-attention in the new windows, connections are established among adjacent windows.}
    \label{fig:patching}
\end{figure}

The CVSSTL and SSTL employ a modified multi-head self attention (MSA) based on shifted windows. 
However, unlike the Swin transformer for optical images, we employ complex-valued attention, layer normalization, and multi-layer perceptron (MLP) layers implemented for 1-D signals. 
Similar to \cite{liu2021swin}, we improve efficiency by computing self-attention within local windows. 
However, simply applying windowed self-attention limits model potential due to the lack of interaction among windows. 
Hence, we employ a shifted window method such that consecutive CVSSTLs or SSTLs employ alternating window partitioning strategies. 
The example shown in Fig. \ref{fig:patching} illustrates the alternating partitioning configuration of two successive modules with a feature dimension of $M = 128$ and window size $W = 64$. 
The first layer employs standard window partitioning starting at the beginning of the signal such that the signal is evenly partitioned into two local windows of 4 patches. 
Alternatively, the second layer computes the windows after shifting the signal by $\lfloor W/2 \rfloor$ samples. 
The shifted window partitioning technique enables connections among adjacent windows in the previous layer and significantly improves modeling capacity \cite{liu2021swin}, similar to forward-backward averaging in signal processing. 

Given an input of size $M \times C$, the CVSSTL or SSTL partitions the signal into $M/W$ non-overlapping windows, resulting in a tensor of shape $M/W \times W \times C$, and computes self-attention independently for each window. 
For each local window feature $\mathbf{Z} \in \mathbb{C}^{W \times C}$, the query, key, and value tensors are computed as
\begin{equation}
    \label{eq:qkv}
    \mathbf{Q} = \mathbf{Z}\mathbf{P}_\mathrm{Q}, \quad \mathbf{K} = \mathbf{Z}\mathbf{P}_\mathrm{K}, \quad \mathbf{V} = \mathbf{Z}\mathbf{P}_\mathrm{V},
\end{equation}
where $\mathbf{P}_\mathrm{Q},\mathbf{P}_\mathrm{K},\mathbf{P}_\mathrm{V} \in \mathbb{C}^{C \times d}$ are the projection tensors and $d$ is the inner dimension of the MSA. 
Then, the self-attention in the local window can be computes as
\begin{equation}
    \label{eq:self_attention}
    \text{Attention}(\mathbf{Q},\mathbf{K},\mathbf{V}) = \mathcal{S}(\mathbf{Q}\mathbf{K}^T/\sqrt{d} + \mathbf{B})\mathbf{V},
\end{equation}
where $\mathcal{S}(\cdot)$ is the softmax function and $B$ is the learnable relative positional encoding. 
However, for complex-valued data, we adopt two major changes to the standard MSA computation \cite{vaswani2017attention}. 
First, the real-valued softmax function is not suited for complex-values, and we introduce a novel complex-valued softmax function that leaves the phase information of the signal unaffected and applies the traditional softmax to the modulus of the signal as 
\begin{equation}
    \label{eq:cvsoftmax}
    \mathcal{S}_\mathbb{C}(\mathbf{X}) = \mathcal{S}_\mathbb{R}(|\mathbf{X}|) \frac{\mathbf{X}}{|\mathbf{X}|},
\end{equation}
where $\mathcal{S}_\mathbb{R}(\cdot)$ is the real-valued softmax function. 
Additionally, for the CVSSTL, we employ a complex-valued $B \in \mathbb{C}^{W \times W}$ for relative positional encoding allowing the learned encoding to impact both the real and imaginary parts of the signal. 
For multi-head attention, we perform the attention computation in \eqref{eq:self_attention} $h$ times and concatenate the results \cite{liang2021swinir}. 

After the MSA, a multi-layer perceptron (MLP) composed of two linear layers separated by a non-linear activation. 
For the CVSSTL, complex-valued linear layers and a complex-valued activation function employed. 
Additionally, we introduce complex-valued layer normalization, which requires careful handling to ensure proper whitening is achieved similarly to complex-valued batch normalization explored in \cite{trabelsi2018deep}. 

The entire process of the CVSSTL or SSTL, detailed in Fig. \ref{fig:model_arch}c, can be expressed as
\begin{equation}
\begin{gathered}
    \label{eq:cvsstl}
    \mathbf{G}_{k-1,i}' = \mathsf{H}_\mathrm{MSA}(\mathsf{H}_\mathrm{LN}(\mathbf{G}_{k-1,i})) + \mathbf{G}_{k-1,i}, \\
    \mathbf{G}_{k,i} = \mathsf{H}_\mathrm{MLP}(\mathsf{H}_\mathrm{LN}(\mathbf{G}_{k-1,i}')) + \mathbf{G}_{k-1,i}',
\end{gathered}
\end{equation}
where $\mathsf{H}_\mathrm{MSA}(\cdot)$ denotes the shifted window MSA similar to that of \cite{liu2021swin,liang2021swinir}, $\mathsf{H}_\mathrm{LN}$ is layer normalization, and $\mathsf{H}_\mathrm{MLP}$ denotes the MLP. 
As illustrated in the following section, the proposed Swin-based technique demonstrates superior modeling capacity compared with cResFreq yielding improved frequency estimation and spectral super-resolution performance. 

\section{Experimental Results}
\label{sec:results}

%In this section we overview the proposed real-valued and complex-valued models and their performance on synthetic and real data. 

\subsection{Experimental Setup and Training Data}
\label{subsec:training}

First, we introduce two models: \textit{SwinFreq} and \textit{CVSwinFreq} that employ real-valued and complex-valued signal Swin modules developed in Section \ref{sec:methods}, respectively. 
For both models, we employ a number of feature channels $C = 32$, inner feature dimension $M = 256$, window size of $W = 16$, number of self-attention heads $h = 8$, $D = 3$ and $B = 4$. 
Following \cite{pan2021complexFrequencyEstimation}, SwinFreq computes the signal modulus after the MF module, whereas CVSwinFreq computes the signal modulus prior to the transposed convolution output layer and employs the complex-valued parametric rectified linear unit (CPReLU) \cite{smith2023complex,smith2023dual_radar}. 
To compete with cResFreq's memory demand and modeling robustness, we set the attention dimension $d = 4$ and $d = 2$ for SwinFreq and CVSwinFreq, respectively, resulting in 249.7k and 260.2k parameters, respectively.
Despite having larger $d$, SwinFreq has the smallest model size as CVSwinFreq requires additional parameters for its complex-valued weight matrices. 
The model hyperparameters are determined empirically, but larger models were explored and found to yield superior yet diminishing performance gains. 

The ground truth training data are generated consistently with prior spectral super-resolution and radar image super-resolution efforts \cite{izacard2021datadriven,pan2021complexFrequencyEstimation,smith2022ffh_vit,vasileiou2022efficient,gao2018enhanced} as a superposition of Gaussian functions centered at the known frequency locations. % as
%\begin{equation}
%    \label{eq:gt_data}
%    \mathbf{Y}_{GT}(f) = \sum_{\ell = 0}^{L-1} \exp \left(-\frac{(f - f_\ell)}{2 \sigma_f^2} \right),
%\end{equation}
%where $f_\ell$ is the frequency of the $\ell$-th component, $\sigma_f$ is the standard deviation of the Gaussian kernel that defines the pulse width, and the corresponding multisinusoidal signal is computed by \eqref{eq:practical_multisinusoidal}. 
Hence, the features input to the network are complex-valued and the ground truth expressions, and model output, are real-valued. 
We find that the min-max norm employed in \cite{smith2023dual_radar} outperforms the modulus normalization technique in \cite{pan2021complexFrequencyEstimation} and, hence, apply a min-max norm to the complex-valued data input to the network. 
During training, we employ $\sigma_f = 0.12/N_{SR}$, the standard deviation of the Gaussian kernel that defines the pulse width, and the additive noise realizations in \eqref{eq:practical_multisinusoidal} are recomputed for each batch making in order to mitigate the model fitting to the noise in the training dataset. 
We construct a training dataset with signal dimension $N = 64$, super-resolved spectral size $N_{SR} = 4096$, additive white Gaussian noise (AWGN) with SNRs between -10 and 40 dB, $L$ are drawn from the discrete uniform distribution between 1 and 10, and $f_\ell$ are sampled uniformly between $[-0.5, 0.5)$ with a minimum spacing of $1/2N_{SR}$. 
As with most transformer-based models \cite{liang2021swinir,liu2021swin,vaswani2017attention,dosovitskiy2020image_ViT,smith2023complex}, we find that a larger training dataset of 500000 training samples yields significant performance improvement. 
%The real-valued loss of the model is given by
%\begin{equation}
%    \mathcal{L} = \sum \left( \hat{\mathbf{Y}} - \mathbf{Y}_{GT}(f) \right)^2,
%\end{equation}
%where $\hat{\mathbf{Y}}$ is the spectrum of the input signal $\mathbf{X}$ estimated by the model. 
We employ the AdamW optimizer \cite{liu2021swin} with an initial learning rate of 0.003 and batch size of 256 and MSE loss. % between the real-valued network output and ground truth data generated using \eqref{eq:gt_data}. 

\begin{figure}[ht]
    \centering
    \includegraphics[width=0.75\textwidth]{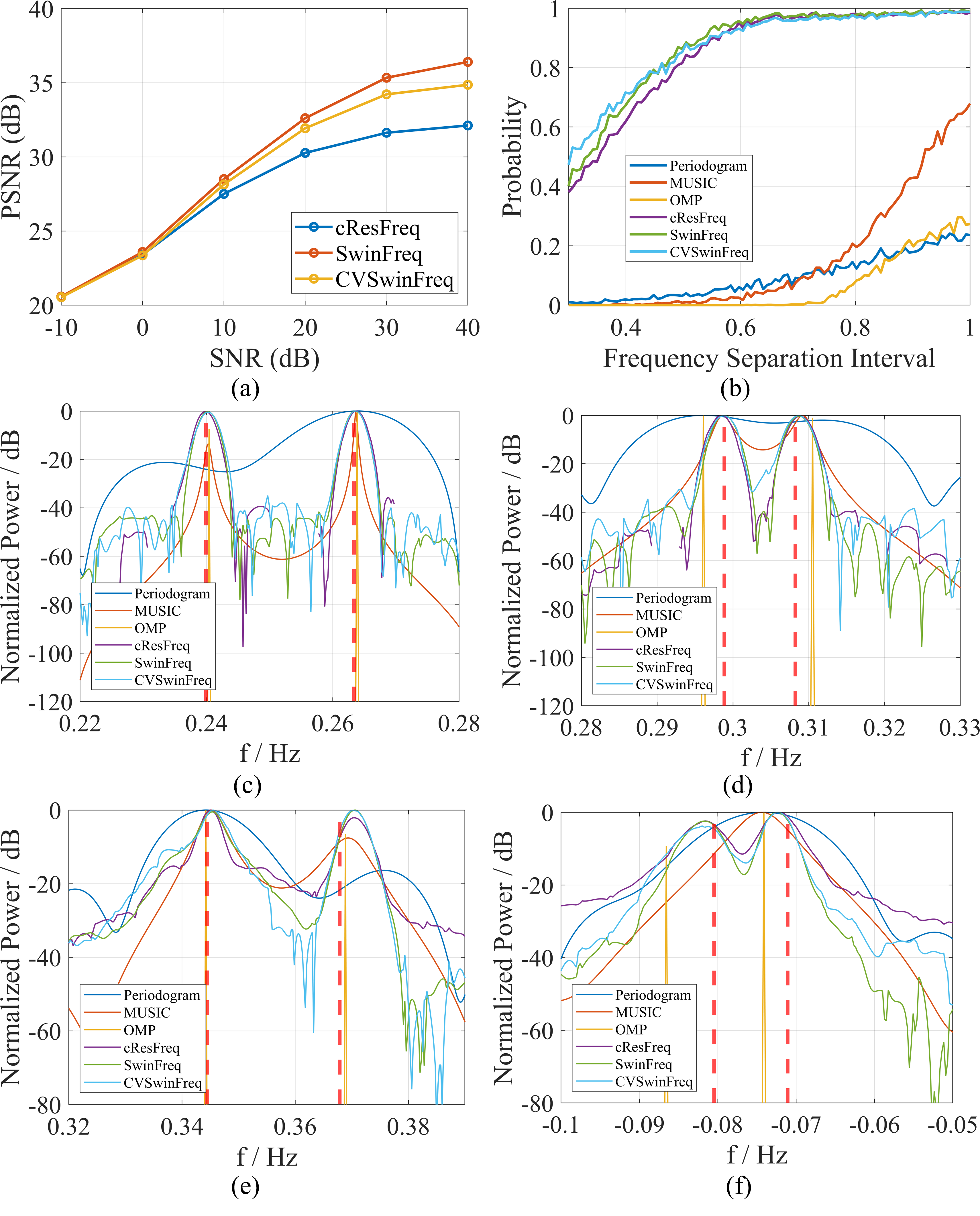}
    \caption{Experimental results on simulated data. (a) Spectral super-resolution performance (PSNR) across SNR. (b) Resolution probability across frequency separation interval ($x/N_{SR}$). Sidelobe comparison for (c)-(d) 20 dB SNR and (e)-(f) 0 dB SNR. Vertical red dashed lines indicate ground truth locations.}
    \label{fig:exp123}
\end{figure}

\subsection{Frequency Estimation Experiments}
\label{subsec:results}

First, we investigate the numerical reconstruction performance of the proposed techniques compared against cResFreq. 
We compare the algorithm performance using the peak signal-to-noise ratio (PSNR) between the ground truth and estimated spectra at different SNR values between -10 and 40 dB, where a larger PSNR indicates a more robust estimate \cite{smith2023dual_radar}, and results are shown in Fig. \ref{fig:exp123}a. 
Particularly at medium to high SNR, both SwinFreq and CVSwinFreq considerably outperform cResFreq with up to a 4.3 dB improvement. 
However, in contrast to other CVNN literature \cite{gao2018enhanced,lee2022complex,trabelsi2018deep}, we note that the performance of SwinFreq exceeds that of CVSwinFreq, despite having fewer parameters. 

Next, we consider the resolution capacity of the periodogram, MUSIC, OMP, and cResFreq in comparison to our algorithms. 
We employ the resolution criterion given two frequency components $f_1$ and $f_2$ as
\begin{equation}
    D = \begin{cases}
        1, \quad \hat{\mathbf{Y}}(f_\mathrm{mid}) < \min(\hat{\mathbf{Y}}(f_1),\hat{\mathbf{Y}}(f_2))/\sqrt{2} \\
        0, \quad \text{else}
    \end{cases},
\end{equation}
where $f_\mathrm{mid}$ is the frequency at the midpoint of $f_1$ and $f_2$ and $\hat{\mathbf{Y}}$ is computed in \eqref{eq:sr_module} \cite{pan2021complexFrequencyEstimation}. 
We perform a Monte Carlo simulation with 20 dB SNR computing the resolution probability over 1000 data samples for each separation interval from $0.3/N_{SR}$ to $1/N_{SR}$ and compute the probability of resolution at each interval. 
As shown in Fig. \ref{fig:exp123}b, SwinFreq and CVSwinFreq demonstrate superior resolution capability than cResFreq, specifically for closely spaced frequencies. 
We also observe that MUSIC and OMP often recover biased estimates that result in incorrect frequency peak locations resulting in inconsistent resolution. 

Additionally, we study the sidelobe reduction robustness of the proposed methods. 
We simulate data at 20 dB and 0 dB SNR with unit amplitude and separations of $0.6/N_{SR}$ and $1.5/N_{SR}$. 
For the 20 dB SNR case, cResFreq achieves similar sidelobe and mainlobe performance to the proposed methods for both closely spaced, Fig. \ref{fig:exp123}c, and distant, Fig. \ref{fig:exp123}d, frequencies. 
However, for lower SNR such as 0 dB, cResFreq demonstrates considerably weaker sidelobe mitigation performance, particularly for closely spaced frequencies, Fig. \ref{fig:exp123}f, resulting in the loss of resolution performance noted in Fig. \ref{fig:exp123}b. 

\begin{figure}[th]
    \centering
    \includegraphics[width=0.75\textwidth]{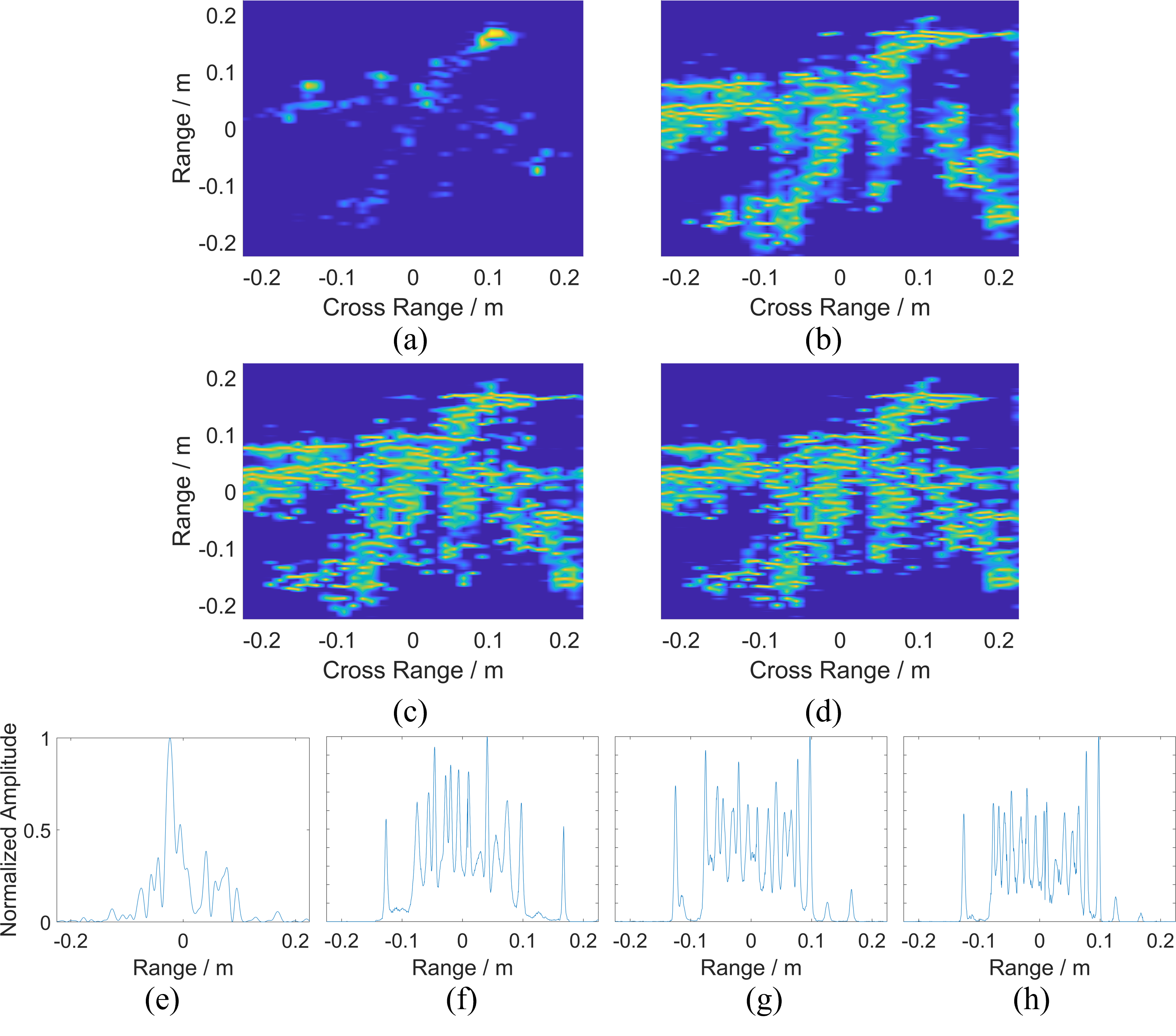}
    \caption{ISAR imaging (a)-(d) and HRRP (e)-(h) results on real data. (a),(e) Windowed periodogram. (b),(f) cResFreq. (c),(g) SwinFreq. (d),(h) CVSwinFreq.}
    \label{fig:exp5}
\end{figure}

Finally, we apply the proposed algorithm to real data in the form of inverse synthetic aperture radar (ISAR) and high-resolution range profile (HRRP) signals, as shown in Fig. \ref{fig:exp5}. 
As detailed in \cite{gao2018enhanced}, ISAR data are collected of an airplane model using a linear frequency modulated (LFM) radar whose bandwidth spans 213.6 GHz to 226.4 GHz.
However, the proposed method is hardware independent. 
The HRRPs shown in Figs. \ref{fig:exp5}e-\ref{fig:exp5}h are extracted from the center of the ISAR images in Figs. \ref{fig:exp5}a-\ref{fig:exp5}d. 
Unsurprisingly, the periodogram extracts weakly-resolved features of the plane, but we see its shape better defined when cResFreq, SwinFreq, or CVSwinFreq are applied. 
We notice SwinFreq and CVSwinFreq show improved definition of the wings compared with cResFreq. 
However, we note the advantage of SwinFreq and CVSwinFreq become when inspecting the HRRPs where the fine details of the plane are more visible and sidelobes are more effectively reduced. 
In both quantitative and qualitative comparisons, the proposed methods demonstrate superior frequency estimation performance compared with existing methods. 

\section{Conclusion}
\label{sec:conclusion}

In this letter, we proposed a novel complex-valued shifted window transformer technique for frequency estimation. 
The proposed method employs a 1-D Swin-based architecture for feature extraction as opposed to convolution yielding superior numerical and empirical performance. 
We introduced real-valued and complex-valued versions of this method, SwinFreq and CVSwinFreq, respectively, that outperform the existing methods in terms of spectral super-resolution robustness, resolution capacity for closely spaced frequency components, and sidelobe reduction at low SNR. 
Notably, SwinFreq outperforms cResFreq in every category while requiring fewer model parameters. 
Experimental results demonstrate the proposed algorithms ability to outperform conventional and data-driven alternatives. 

\bibliography{mega_bib}
\bibliographystyle{IEEEtran}\end{document}